\documentclass[aps,prd,groupedaddress,amsmath,amssymb,amsfonts,nofootinbib,preprint]{revtex4-1}

\usepackage{amssymb,amsfonts}
\usepackage{mathrsfs}
\usepackage{tikz}
\DeclareMathAlphabet{\mathpzc}{OT1}{pzc}{m}{it}
\usetikzlibrary{patterns}

\newcommand{\w}{\frac{\bar{p}}{\bar{\rho}}}

\begin{document}

\bibliographystyle{plain}

\title{The Cosmological Memory Effect}

\author{Alexander Tolish}
\email{tolish@uchicago.edu}
\author{Robert M. Wald}
\email{rmwa@uchicago.edu}
\affiliation{Enrico Fermi Institute and Department of Physics \\
  The University of Chicago \\
  5640 S. Ellis Ave., Chicago, IL 60637, U.S.A.}

\begin{abstract}
The ``memory effect'' is the permanent change in the relative separation of test particles resulting from the passage of gravitational radiation. We investigate the memory effect for a general, spatially flat FLRW cosmology by considering the radiation associated with emission events involving particle-like sources. We find that if the resulting perturbation is decomposed into scalar, vector, and tensor parts, only the tensor part contributes to memory. Furthermore, the tensor contribution to memory depends only on the cosmological scale factor at the source and observation events, not on the detailed expansion history of the universe. In particular, for sources at the same luminosity distance, the memory effect in a spatially flat FLRW spacetime is enhanced over the Minkowski case by a factor of $(1 + z)$.
\end{abstract}

\maketitle

\section{Introduction}
\label{intro}

The passage of gravitational radiation through a configuration of test particles that make up a gravitational wave detector can induce a permanent change in the relative separation of the test particles, a phenomenon which has come to be known as the \textit{gravitational wave memory effect}. The memory effect on a flat background was first recognized in the linear regime for sources in non-relativistic motion by Zel'dovich and Ponarev \cite{Zeldovich}. Afterwards, Christodoulou \cite{Christodoulou} discovered that there could be additional contributions to memory arising from the nonlinearity of the Einstein equation and associated with the Bondi flux of the gravitational waves to null infinity. Shortly thereafter, it was argued that the \textit{nonlinear} memory of Christodoulou could be interpreted as corresponding to a \textit{linear} memory caused by the effective stress-energy associated with the primary gravitational radiation \cite{Thorne}, \cite{WisemanWill}. 

More recently, it has been found to be useful to make a distinction between \textit{ordinary} and \textit{null} memory in the linearized gravity context \cite{BG1},\cite{BG2}: Ordinary memory is caused by massive matter sources and null memory is caused by null matter sources. The gravitational radiation emitted from a source event involving both massive and null matter will induce both ordinary and null memory \cite{BG1},\cite{BG2},\cite{TBGW}. Neither the ordinary nor null memory should be viewed as a tidal effect with a Newtonian analogue but rather as a byproduct of the gravitational radiation emitted from a burst-type event \cite{TBGW},\cite{TW}. 

The above work has concerned itself with memory on an asymptotically flat spacetime. If a spacetime is not asymptotically flat, it is not clear what ``memory'' should mean, even if we treat the gravitational radiation as a perturbation off of a background metric. Memory has been defined in terms of the net change in the separation of test particles, but this could include motion due to the background curvature rather than the radiation. For example, in cosmological Friedmann-Lema\^\i tre-Robertson-Walker (FLRW) cosmology, the proper separation of FLRW observers will change with time. Furthermore, if there is no notion of null infinity, it is not clear where we should put our detector so that it will be exposed to radiation but isolated from other non-radiative gravitational forces.

Recently, Bieri, Garfinkle and Yau \cite{BGY}, Kehagias and Riotto \cite{Kehagias}, and Chu \cite{Chu1},\cite{Chu2} have considered the memory effect in FLRW spacetimes. However, their choice of methods for resolving the above difficulties have so far limited the applicability of their results. Bieri and Garfinkle consider Weyl tensor perturbations, so their analysis is greatly simplified by working \textit{in vacuo}, and, consequently, they have considered only memory in a background vacuum de Sitter universe. Meanwhile, Kehagias and Riotto's use of BMS transformations depends on the existence of a null infinity for the FLRW spacetime, so they restrict consideration to a decelerating universe without a cosmological constant. Furthermore, both groups consider only null memory. While Chu does not make either of these specific restrictions, his preferred definition of memory does not distinguish test particle motion due to radiative and non-radiative gravity.

In this paper, we will investigate the total memory effect---both ordinary and null---in a general, spatially flat, FLRW spacetime. We will simplify the problem not by limiting the cosmological models under consideration, but rather the kinds of radiation sources. Specifically, we will consider---in the context of linear\footnote{Particle-like sources in Einsten's equation do not make sense outside of the context of linear perturbation theory \cite{GT}.} perturbation theory off of an FLRW background---only point-particle sources, \textit{i.e.}, sources whose stress-energy is confined by Dirac delta functions to worldlines that meet at a vertex, which we will call the \textit{source event}. We previously considered such sources in Minkowski spacetime \cite{TBGW}, \cite{TW}, and found that the retarded solution gives rise to a contribution to the curvature tensor of the form of the derivative of a delta function in retarded time. Upon integration of the geodesic deviation equation, this derivative of a delta function in curvature gives rise to a well defined memory effect---i.e., a change in the separation of test particles coincident in time with the passage of the radiation from the source event---that includes both the ordinary and null memory  \cite{TBGW}, \cite{TW}. Thus, for the idealized sources that we consider, the memory effect can be associated with the presence of a derivative of a delta function in the Riemann curvature of the retarded solution arising from the source event. This enables us to distinguish the memory effect induced by the passage of radiation from non-radiative gravitational effects, and it does not require us to take limits to null infinity. It thereby yields a well defined notion of the memory effect that is applicable to linearized perturbations about arbitrary background spacetimes.

Another advantage to considering the above idealized particle sources, where the radiation emerges from a single ``source event'' in the background spacetime, is that---since all spacetimes are ``locally flat'' on sufficiently small scales---there is a well defined notion of having the ``same source'' in different spacetimes. Similarly, there is a well defined notion of having the ``same detector''---i.e., geodesic test particles initially at rest and with small separation---in different spacetimes. Thus, we can compare the memory effect in two different spacetimes provided only that we specify the location and ``rest frame'' of both the source and the detector in the two spacetimes. Since any spatially flat FLRW spacetime is conformal to Minkowski spacetime, it is particularly useful to state our results by comparing the memory effect in FLRW spacetime to that in Minkowski spacetime. Stated in this manner, the main result of our paper is as follows: 

\textit{Consider a spatially flat FLRW solution to Einstein's equation with arbitrary fluid matter, and with a given particle source perturbation, as described above. Now place the same source and detector in Minkowski spacetime such that the  source and detector are at rest with respect to each other and the source is at a distance, $d$, in Minkowski spacetime equal to the luminosity distance, $d_L$, in the FLRW spacetime. Then the memory effect in the FLRW spacetime is enhanced over the Minkowski value by a factor of $(1+z)$, where $z$ denotes the redshift factor between the source and observer/detector.}

This result applies to both ordinary and null memory. It is in agreement with the results obtained for null memory in the special cases considered in \cite{BGY} and \cite{Kehagias}. 
Note that in a spatially flat FLRW spacetime, the luminosity distance $d_L$ and the angular diameter distance, $d_A$, are related by
\begin{equation}
d_A = d_L/(1+z)^2 \label{distance}
\end{equation}
It follows that for a Minkowski source at $d = d_A$, the memory effect in the FLRW spacetime will be decreased from the Minkowski value by a factor of $(1+z)^{-1}$. Finally, it should be noted that although the memory effect in a spatially flat FLRW spacetime is related to the Minkowski memory effect in this simple way, the waveforms will be different; in particular, there will be ``tail effects'' in the FLRW spacetime.

Although our analysis is restricted to the context of linear perturbation theory with idealized particle sources, we expect that our main result stated above should be valid completely generally for any sources whose spatial and time variation scales are small compared with the Hubble scale. Indeed, the main difficulty in generalizing our results to non-particle-like sources and to the nonlinear regime would be to give a precise definition of ``memory'' outside of the context we consider. Thus, if one wishes to compute the memory effect resulting from, say, the coalescence of two black holes in a distant galaxy in a spatially flat FLRW spacetime, it should suffice to compute the memory effect arising from a similar coalescence in an asymptotically flat spacetime and then use the above correspondence. However, we shall not attempt to formulate or prove such a generalization here.

We shall begin in section~\ref{sec:idealized} by describing the particle sources that we shall consider and characterizing the memory effect for perturbations of arbitrary curved spacetimes. In section~\ref{sec:perttheory}, we analyze linearized perturbations off of spatially flat FLRW spacetimes with such particle sources. In section~\ref{sec:retgrav}, we consider the tensor mode contribution to memory. We show that only the ``light cone portion'' of the retarded Green's function will contribute to memory, and that its contribution can be related in a simple way to the memory caused by similar sources in a flat spacetime. In the Appendix, we show that the scalar modes do not contribute to memory.

Latin indices from the early alphabet ($a,b, \dots$) denote abstract spacetime indices. Greek indices ($\mu, \nu, \dots$) denote spacetime components of tensors, whereas Latin indices from the mid-alphabet ($i,j, \dots$) denote spatial components. 

\section{Particle Sources}
\label{sec:idealized}

In asymptotically flat spacetimes, the memory effect can be characterized in a precise manner by considering a detector composed of test particles near null infinity. Radiation effects fall off as $1/r$, whereas Newtonian tidal effects fall off as $1/r^3$, so by considering only the $O(1/r)$ effects on the test particles, we can distinguish between effects produced by gravitational radiation and all other gravitational effects. However, in a non-asymptotically-flat spacetime, it is not clear how even to define memory, since there is no obvious way to make a clean distinction between effects due to ``radiation'' as compared with other tidal gravitational effects.

In our previous investigation of the memory effect in linearized gravity \cite{TBGW},\cite{TW}, we considered the idealized process of the instantaneous decay of a massive particle into two other particles. We found that the retarded solution to the linearized Einstein equation with such a source has the property that the $O(1/r)$ part of the curvature has the form of a derivative of a delta-function of retarded time at the retarded time of the decay event. Integration of the geodesic deviation equation then shows that the $O(1/r)$ effect on test particles is to produce a sharp step function in their relative separation. Thus, for this kind of idealized source, the memory effect can be characterized by the presence of a derivative of a delta-function in the linearized curvature and a corresponding step function behavior in the relative separation of test particles\footnote{Of course, if one were to consider a less idealized source with a smoothed out energy-momentum tensor, then the Riemann tensor also will be smoothed out, and the relative separation of the test particles will not undergo a sharp, sudden change in separation; rather, separation of the particles that results in a memory effect would occur continuously on the same timescale as that of the event itself.}.

For our present purposes, the main advantage of considering sources consisting of particles undergoing instantaneous interactions is that the characterization of the memory effect in terms of derivative of a delta-function behavior of the curvature holds at all distances from the interaction event, i.e., one does not need to go to null infinity to extract this characterization of the memory effect. This characterization may therefore be imported straightforwardly to other spacetimes. Thus, in this paper, we shall restrict consideration to linearized gravity off of a smooth background spacetime, with a linearized perturbation sourced by (massive or massless) point particles. The interactions of the particles will be modeled by having their worldlines intersect (and, possibly, begin or end) at a single event, $q$, in spacetime as illustrated in figure 1. Conservation of stress-energy then requires that (i) the particle worldlines are geodesics \cite{GrallaWald} away from $q$, and (ii) total $4$-momentum is conserved at $q$. ``Memory'' will then be characterized by the presence of a derivative of a delta-function in the curvature of the retarded solution with this source. This characterization does not require that the detector be placed near ``infinity.''

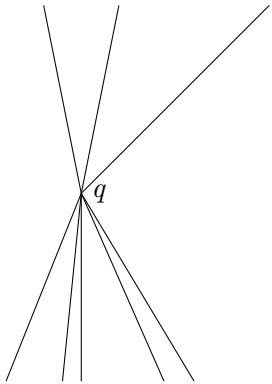
\begin{figure}[ht]
\centering
\begin{tikzpicture}
\draw (1.5,0)  -- (2.5,2.5);
\draw (2.25,0) -- (2.5,2.5);
\draw (2.5,0)  -- (2.5,2.5);
\draw (3.6,0)  -- (2.5,2.5);
\draw (4,0)    -- (2.5,2.5);
\draw (2.5,2.5)-- (2,5);
\draw (2.5,2.5)-- (3,5);
\draw (2.5,2.5)-- (5,5);
\node [right] at (2.5,2.5) {\textit{q}}; 
\end{tikzpicture}
\caption{A spacetime diagram of the sort of gravitational wave source we will consider. Here $5$ point particles enter a single ``source event'' $q$, and $3$ emerge. The worldlines of the incoming and outgoing particles must be timelike or null geodesics.}
\end{figure}

To specify more precisely the type of source we consider, we assume that local coordinates 
$(t,\mathbf{x})$ have been introduced in a neighborhood of $q$ so that $\nabla t$ is past-directed timelike and so that the event $q$ is labeled by $t=\mathbf{x} = 0$. The worldline, $\gamma$, of each incoming massive particle must be a timelike geodesic \cite{GrallaWald} with endpoint at $q$. We can parametrize $\gamma$ by $t$ and specify it by giving $\mathbf{x}(t) = \mathbf{z}(t)$, where $\mathbf{z}(0) = \mathbf{0}$. The stress-energy of each incoming massive particle then takes the form
\begin{equation}
T^{(M, {\rm in})}_{ab} = m u_a u_b \, \delta^{(3)}\left(\mathbf{x} - \mathbf{z}(t)\right) \frac{1}{\sqrt{-g}} \frac{d\tau}{dt} \Theta(-t) \, .
\label{particle}
\end{equation}
Here $u^{a}$ is the unit tangent ($4$-velocity) to $\gamma$, $\tau$ is the proper time along $\gamma$, $\Theta$ is the Heaviside step function, and $\delta^{(3)}$ is the ``coordinate delta-function,'' i.e., $\int \delta^{(3)}(\mathbf{x} - \mathbf{z}(t)) d^3 \mathbf{x} = 1$. Each incoming massless particle moves on a null geodesic, $\alpha$, given by $\mathbf{x}(t) = \mathbf{y}(t)$, with $\mathbf{y}(0) = \mathbf{0}$. The stress-energy of each incoming massless particle takes the form
\begin{equation}
T^{(N, {\rm in})}_{ab} = k_a k_b \, \delta^{(3)}\left(\mathbf{x} - \mathbf{y}(t)\right) \frac{1}{\sqrt{-g}} \frac{d\lambda}{dt} \Theta(-t) \, .
\label{nullparticle}
\end{equation}
Here, $\lambda$ is an affine parameter of $\alpha$ and $k^a$ is the corresponding tangent, with the scaling of $\lambda$ chosen so that \eqref{nullparticle} holds. 
 
The stress-energy of each of the outgoing massive and massless particles takes the form of \eqref{particle} and \eqref{nullparticle} except that $\Theta(-t)$ is replaced by $\Theta(t)$. The total stress-energy of the particle sources we consider takes the form
\begin{equation}
T^{(P)}_{ab} = \sum_{l, {\rm in}} T^{(M,l)}_{ab} + \sum_{n, {\rm in}} T^{(N,n)}_{ab}  + \sum_{l', {\rm out}} T^{(M, l')}_{ab} + \sum_{n', {\rm out}} T^{(N, n')}_{ab} \, ,
\label{totalT}
\end{equation}
where each $T^{(M,l)}_{ab}$ takes the form of \eqref{particle},  each $T^{(N,n)}_{ab}$ takes the form of \eqref{nullparticle}, and
each $T^{(M, l')}_{ab}$ and $T^{(N,n')}_{ab}$ also take these forms with $\Theta(t)$ replaced by $\Theta(-t)$. Conservation of stress-energy, $\nabla^a T_{ab} = 0$, holds in the distributional sense away from $q$ by virtue of the fact that each particle moves on a geodesic \cite{GrallaWald}. Conservation of stress-energy will hold at $q$ if and only if we have at $q$
\begin{equation}
\sum_{l, {\rm in}} m^{(l)}u^{(l)}_a+ \sum_{n, {\rm in}} k^{(n)}_a =  \sum_{l', {\rm out}} m^{(l')}u^{(l')}_a + \sum_{n', {\rm out}} k^{(n')}_a \, ,
\label{cons}
\end{equation}

Equation \eqref{totalT} with condition \eqref{cons} defines the sources that we consider in this paper. We are interested in the solution to the linearized Einstein equation ``produced by such a source'' in an arbitrary background spacetime. For hyperbolic equations, what we mean by ``produced by a source'' is the solution obtained by convolving the source with the retarded Green's function. In general spacetimes, issues of convergence of the retarded solution will arise from the contribution of the sources at arbitrarily early times\footnote{Even in Minkowski spacetime, the contribution of null sources at arbitrarily early times does not converge to a distribution \cite{TW}.}. However, we are not interested in such issues of convergence here, but rather the effects arising from near the source event $q$. Thus, we shall simply consider the contributions to the retarded Green's function integral arising from the above particle sources in a small neighborhood of $q$. The memory effect will then be identified with the presence of a derivative of a delta-function in the curvature ``produced'' by these particle sources near event $q$.

\section{Perturbation Theory in Cosmological Spacetimes}\label{sec:perttheory}

We now wish to consider linearized perturbations off of a spatially flat FLRW background,
\begin{equation}
ds^2= - d \tau^2 +a^2(\tau)(dx^2 + dy^2 + dz^2).
\end{equation}
As usual, it is convenient to introduce conformal time $d\eta=d\tau/a$ so that the background FLRW metric takes the manifestly conformally flat form
\begin{equation}
ds^2 =a^2(\eta)\left(- d \eta^2 + dx^2 + dy^2 + dz^2\right).
\label{concoor}
\end{equation}
Throughout the rest of the paper, ``$0$'' and ``$i$'' (i.e., spatial) indices will denote components of tenors with respect to these
coordinates, and an overdot will denote a derivative with respect to $\eta$. We will write $\partial^i=\delta^{ij}\partial_j$ and $\nabla^2 = \partial^i\partial_i = \delta^{ij} \partial_i \partial_j$, i.e., $\nabla^2$ is the Laplacian with respect to the spatial metric $\delta_{ij}$ given by $dx^2 + dy^2 +dz^2$.

We assume that Einstein's equation holds (possibly with a cosmological constant $\Lambda$) and that the matter stress-energy---apart from
the particle matter that we will add as a perturbation---is that of a perfect fluid
\begin{equation}
T^{(F)}_{ab}=(\rho+p)u_a u_b+pg_{ab} \, ,
\label{fluidT}
\end{equation}
with $4$-velocity $u^a$, density $\rho$ and pressure $p$. 
The fluid is assumed to be described by a one-parameter (``barotropic'') equation of state $p=p(\rho)$. The density and pressure are perturbations away from homogeneous background values $\bar{\rho}$ and $\bar{p}$ which satisfy the Friedmann equations
\begin{gather}
\left(\frac{1}{a}\frac{da}{d\tau}\right)^2=\frac{1}{3}\left(8\pi\bar{\rho}+\Lambda\right) \, ,\label{Friedmann1}\\
\frac{1}{a}\frac{d^2a}{d\tau^2}=\frac{1}{3}\left(-4\pi(\bar{\rho}+3\bar{p})+\Lambda\right) \, . \label{Friedmann2}
\end{gather}

We write the perturbed metric as
\begin{equation}
g_{ab}= \bar{g}_{ab}+a^2h_{ab}
\end{equation}
where $\bar{g}_{ab}$ denotes the background FLRW metric. The perturbed fluid is described by $\delta u^\mu$, $\delta \rho$, and $\delta p = c_s^2 \delta \rho$, where
\begin{equation}
c_s^2=\frac{d p}{d\rho} \label{speedofsound} \, .
\end{equation}
We wish to consider the metric perturbation resulting
from the presence of a particle stress-energy of the form \eqref{totalT}. The particle sources are assumed to have no
direct interaction with the fluid present in the FLRW background; the particle stress-energy is separately conserved. 
However, since the particles affect the perturbed metric, they automatically affect the fluid (even at the linearized level),
so the fluid perturbations cannot be ignored. 

Analysis of the perturbations is most easily done using the gauge-invariant methods of Bardeen \cite{Bardeen} with modifications by Durrer \cite{Durrer1},\cite{Durrer2} allowing for additional forms of matter perturbations\footnote{Both Bardeen and Durrer allow for general stress-energies with non-fluid properties like anisotropic pressures. However, as we have discussed in section \ref{sec:idealized}, we want our perturbed fluid and particles to interact only gravitationally, which means that the stress-energies of the fluid and the particles must be conserved independently. Bardeen does not discuss this scenario, but it corresponds to Durrer's notion of cosmological seeds.}. These methods rely on decomposing the metric, fluid stress-energy, and particle stress-energy perturbations into scalar, vector, and tensor parts, and working with gauge invariant quantities in each sector. 
We can decompose a general symmetric tensor field $X_{ab}$ on spacetime into its scalar, vector, and tensor parts by writing its coordinate components as
\begin{equation}
X_{\mu\nu}=\left(\begin{array}{c|c}
\varphi&\partial_i\chi\;\;+\;\;\xi_i\\
\hline
{\begin{array}{c}\partial_i\chi\\+\\ \xi_i\end{array}}&{\begin{array}{c}\psi\delta_{ij}+(\partial_i\partial_j-\frac{1}{3}\delta_{ij}\nabla^2)\omega\\+\partial_{(i}\zeta_{j)}+\mathscr{X}_{ij}\end{array}}
\end{array}\right) \, ,
\label{decomp}
\end{equation}
where the scalar parts are given by\footnote{If the spatial slices have topology $\bf{R}^3$, we need to impose boundary conditions at infinity in order to get a unique solution to the Poisson equations for $\chi$ and $\omega$ (and $\zeta_i$ below), which, in turn, may put restrictions on the asymptotic behavior of $X_{ab}$. However, as we are ultimately interested in singular behavior of the perturbations, it does not matter what solutions of the Poisson equations we choose. For convenience, we shall assume that the spatial slices have the topology of three-tori, with the dimensions of the tori being much larger than the dimensions of the physical problem. The solutions are then unique up to the addition of constants, which do not affect the decomposition.}
\begin{align}
\varphi&=X_{00}\nonumber\\
\nabla^2\chi&=\partial^iX_{0i}\nonumber\\
\psi&=\frac{1}{3}\delta^{ij}X_{ij}\nonumber\\
\nabla^2\nabla^2\omega&=\frac{3}{2}\left(\partial^i\partial^j-\frac{1}{3}\delta^{ij}\nabla^2\right)X_{ij} \, ,\label{scalar}
\end{align}
the vector parts are given by
\begin{align}
\xi_i&=X_{0i}-\partial_i\chi\nonumber\\
\nabla^2\zeta_i&=2\left(\partial^jX_{ij}-\partial_i\psi-\frac{2}{3}\nabla^2\partial_i\omega\right) \, ,\label{vector}
\end{align}
and the tensor part is
\begin{equation}
\mathscr{X}_{ij} =X_{ij}-\psi\delta_{ij}-\left(\partial_i\partial_j-\frac{1}{3}\delta_{ij}\nabla^2\right)\omega-\partial_{(i}\zeta_{j)} \, . \label{svt}
\end{equation}

If the metric perturbation is written in this way,
\begin{equation}
h_{\mu\nu}=\left(\begin{array}{c|c}
\varphi^{(h)}&\partial_i\chi^{(h)}\;\;+\;\;\xi^{(h)}_i\\
\hline
{\begin{array}{c}\partial_i\chi^{(h)}\\+\\ \xi^{(h)}_i\end{array}}&{\begin{array}{c}\psi^{(h)}\delta_{ij}+(\partial_i\partial_j-\frac{1}{3}\delta_{ij}\nabla^2)\omega^{(h)}\\+\partial_{(i}\zeta^{(h)}_{j)}+\mathpzc{h}_{ij}\end{array}}
\end{array}\right),
\end{equation}
then
\begin{align}
\Phi&=\varphi^{(h)}+2\dot{\chi}^{(h)}+2\frac{\dot{a}}{a}\chi^{(h)}+\ddot{\omega}^{(h)}+\frac{\dot{a}}{a}\omega^{(h)}\\
\Psi&=\psi^{(h)}+2\frac{\dot{a}}{a}\chi^{(h)}-\frac{1}{3}\nabla^2\omega^{(h)}-\frac{\dot{a}}{a}\dot{\omega}^{(h)}
\end{align}
are gauge-invariant scalar quantities, whereas
\begin{equation}
\Xi_i=\xi^{(h)}_i-\dot{\zeta}^{(h)}_i
\end{equation}
is a gauge-invariant vector quantity, and
$\mathpzc{h}_{ij}$ is a gauge-invariant tensor quantity. The above two scalar fields, $\Phi$ and $\Psi$, one transverse three-vector field, $\Xi_i$, and one transverse-traceless three-tensor field, $\mathpzc{h}_{ij}$, contain all of the physical (non-gauge) information concerning the metric perturbation. 

The stress-energy tensor of the particles \eqref{totalT} can also be decomposed in this way:
\begin{equation}
T^{(P)}_{\mu\nu}=\left(\begin{array}{c|c}
\varphi^{(P)}&\partial_i\chi^{(P)}\;\;+\;\;\xi^{(P)}_i\\
\hline
{\begin{array}{c}\partial_i\chi^{(P)}\\+\\ \xi^{(P)}_i\end{array}}&{\begin{array}{c}\psi^{(P)}\delta_{ij}+(\partial_i\partial_j-\frac{1}{3}\delta_{ij}\nabla^2)\omega^{(P)}\\+\partial_{(i}\zeta^{(P)}_{j)}+\mathscr{T}_{ij}\end{array}}
\end{array}\right).\label{particleSE}
\end{equation}
Because there is no ``background'' particle stress-energy, each of the individual component fields $\varphi^{(P)},\chi^{(P)},$ etc. are already gauge invariant to first order. These quantities are related to $T^{(P)}_{\mu\nu}$ by eqs.(\ref{scalar})-(\ref{svt}). Since $T^{(P)}_{\mu\nu}$ is distributional, these quantities will also be distributional.

We can also find gauge-invariant combinations of the perturbed stress-energy $\delta T^{(F)}_{\mu \nu}$ of the fluid, \eqref{fluidT}. We define
\begin{equation}
\delta_\rho= \frac{\rho-\bar{\rho}}{\bar{\rho}} \, .
\end{equation}
We decompose the perturbed $4$-velocity as
\begin{equation}
\delta u^\mu=\frac{1}{a}\left(\begin{array}{c}
\delta u^0\\
\partial^i v+v^i\end{array}\right)
\end{equation}
with $\partial_i v^i = 0$, and we remind the reader that $\partial^i v = \delta^{ij} \partial_j v$.
Note that the quantity $\delta u^0$ is not independent, since it is fixed by the normalization condition $g_{ab}u^au^b=-1$. In terms of these quantities and the perturbed metric, we can obtain the following gauge-invariant fluid variables:
\begin{gather}
V=v+\frac{1}{2}\dot{\omega}^{(h)}\\
A=\delta_\rho+3\left(1+\frac{\bar{p}}{\bar{\rho}}\right)\left(\frac{1}{2}\left(\psi^{(h)}-\frac{1}{3}\nabla^2\omega^{(h)}\right)-\frac{\dot{a}}{a}V-\Phi\right)\\
W_i=\delta_{ij}v^j+\frac{1}{2}\dot{\xi}^{(h)}_i.
\end{gather}
The fields $V$, $A$, and $W_i$ thus provide us, respectively, with gauge-invariant measures of fluid's peculiar velocity with respect to the Hubble flow, its perturbed density, and its vorticity.

The linearized Einstein equation decomposes into decoupled sets of equations involving the scalar, vector, and tensor parts of the perturbations. These equations can be written entirely in terms of the gauge-invariant quantities introduced above. The scalar equations are
\begin{gather}
\nabla^2\Psi-3\frac{\dot{a}}{a}\left(\dot{\Psi}+\frac{\dot{a}}{a}\Phi\right)  = -8\pi\left(a^2\bar{\rho}A-3a\dot{a}(\bar{\rho}+\bar{p})V+\varphi^{(P)}\right)\label{EFE_1}\\
\partial_i\left(\dot{\Psi}+\frac{\dot{a}}{a}\Phi\right)  = -8\pi\left(a^2(\bar{\rho}+\bar{p})\partial_iV-\partial_i\chi^{(P)}\right)\label{EFE_2}\\
\partial_i\partial_j\left(\Psi-\Phi\right)  = -16\pi\partial_i\partial_j\omega^{(P)}\label{EFE_3}
\end{gather}
\begin{multline}
\ddot{\Psi}+2\frac{\dot{a}}{a}\dot{\Psi}+\frac{\dot{a}}{a}\dot{\Phi}+\left(2\frac{\ddot{a}}{a}-\left(\frac{\dot{a}}{a}\right)^2\right)\Phi+\frac{2}{3}\nabla^2\left(\Psi-\Phi\right)\\=-16\pi\left(a^2c_s^2(\bar{\rho})A-3(\bar{\rho}+\bar{\phi})\frac{\dot{a}}{a}V+\psi^{(P)}\right) \, .\label{EFE_4}
\end{multline}
The vector equations are
\begin{gather}
\nabla^2\Xi_i=-8\pi\left(2a^2(\bar{\rho}+\bar{p})W_i-\xi^{(P)}_i\right)\label{EFE_5}\\
\partial_{(i}\dot{\Xi}_{j)}+2\frac{\dot{a}}{a}\partial_{(i}\Xi_{j)}=-8\pi\partial_{(i}\zeta^{(P)}_{j)} \, .\label{EFE_6}
\end{gather}
Finally, the tensor equation is
\begin{equation}
-\ddot{\mathpzc{h}}_{ij}-2\frac{\dot{a}}{a}\dot{\mathpzc{h}}_{ij}+\nabla^2\mathpzc{h}_{ij} =-16\pi \mathscr{T}_{ij}\label{EFE_7} \, .
\end{equation}

If the various perturbation fields do not grow in an unbounded fashion at large distances, the unique solutions to \eqref{EFE_2} and \eqref{EFE_3} are
\begin{gather}
\dot{\Psi}+\frac{\dot{a}}{a}\Phi=-8\pi\left(a^2(\bar{\rho}+\bar{p})V-\chi^{(P)}\right)\label{EFE2}\\
\Psi-\Phi=-16\pi\omega^{(P)} \, .\label{EFE3}
\end{gather}
Equations \eqref{EFE_1} and \eqref{EFE_4} then simplify to
\begin{equation}
\nabla^2\Psi=-8\pi\left(a^2\bar{\rho}A+\varphi^{(P)}+3\chi^{(P)}\right)\label{EFE1}
\end{equation}
\begin{multline}
\ddot{\Psi}+2\frac{\dot{a}}{a}\dot{\Psi}+\frac{\dot{a}}{a}\dot{\Phi}+\left(2\frac{\ddot{a}}{a}-\left(\frac{\dot{a}}{a}\right)^2\right)\Phi\\=-16\pi\left(a^2c_s^2\left(\bar{\rho}A-3(\bar{\rho}+\bar{p})\frac{\dot{a}}{a}V\right)+\psi^{(P)}-\frac{2}{3}\nabla^2\omega^{(P)}\right) \, .\label{EFE4}
\end{multline}
Similarly, \eqref{EFE_6} implies that 
\begin{equation}
\dot{\Xi}_{i}+2\frac{\dot{a}}{a}\Xi_{i}=-8\pi\zeta^{(P)}_{i} \, .\label{EFE6}
\end{equation}
The full set of Einstein's equation thus reduces to \eqref{EFE2}-\eqref{EFE6} together with \eqref{EFE_5} and \eqref{EFE_7}.

The perturbed conservation of stress energy for the fluid, $\delta [\nabla^\mu T^{(F)}_{\mu \nu}] = 0$, yields the scalar equations
\begin{gather}
\dot{V}+\frac{\dot{a}}{a}V=-\frac{c_s^2\bar{\rho}E}{\bar{\rho}+\bar{p}}-\frac{1}{2}\Phi\label{con1}\\
\dot{A}-3\w\frac{\dot{a}}{a}A=-\left(1+\w\right)\left(\nabla^2V-3\chi^{(P)}\right)\label{con2}
\end{gather}
as well as the vector equation
\begin{equation}
\dot{W}_i-3\frac{\dot{a}}{a}c_s^2W_i=0\label{con3} \, .
\end{equation}
Similarly, conservation of stress-energy for the particles can also be expressed in terms of the fields \eqref{particleSE} as
\begin{gather}
\dot{\varphi}^{(P)}+\frac{\dot{a}}{a}\varphi^{(P)}-\nabla^2\chi^{(P)}+3\frac{\dot{a}}{a}\psi^{(P)}=0\label{con4}\\
\dot{\chi}^{(P)}+2\frac{\dot{a}}{a}\chi^{(P)}-\psi^{(P)}-\frac{2}{3}\nabla^2\omega^{(P)}=0\label{con5}\\
\dot{\xi}^{(P)}_i+2\frac{\dot{a}}{a}\xi^{(P)}_i-\frac{1}{2}\nabla^2\zeta^{(P)}_i=0\label{con6}.
\end{gather}

A very useful equation for $A$ can be derived as follows \cite{Durrer1}: We differentiate \eqref{con2} with respect to $\eta$, and substitute from \eqref{con1} to eliminate $\dot{V}$. Then we use \eqref{con2} to eliminate $\nabla^2 V$, and we use \eqref{EFE3} and \eqref{EFE1} to eliminate $\nabla^2\Phi$. Finally, we use \eqref{con5} to eliminate $\dot{\chi}^{(P)}$.
We thereby obtain a wave equation for $A$ with particle sources,
\begin{multline}
-\ddot{A}-\frac{\dot{a}}{a}\left(1+3c_s^2-6\w\right)\dot{A}+c_s^2\nabla^2 A+3\left(\frac{\ddot{a}}{a}\w-3\left(\frac{\dot{a}}{a}\right)^2\left(c_s^2-\w\right)+a^2\left(1+\w\right)\frac{4\pi}{3}\bar{\rho}\right)A\\=-4\pi a^2\left(1+\w\right)\left(\varphi^{(P)}+3\psi^{(P)}\right).\label{soundwave}
\end{multline}
Physically, this equation describes the propagation of sound waves in the fluid. Although there is no direct coupling between the particles and the fluid, there are ``particle source terms'' in \eqref{soundwave} resulting from the gravitational interactions between the particles and the fluid.

As previously explained, the memory effect will be identified with the presence of a derivative of a delta function in the curvature. The curvature is given by an expression involving at most $2$ derivatives of the metric variables. In particular, in the Newtonian gauge, the perturbation to the electric components of the Riemann tensor is \cite{Durrer3}
\begin{multline}
\delta R_{i00}^{\quad j}=-\frac{1}{2}\Bigg(\left(\partial_i\partial_k-\frac{1}{3}\delta_{ik}\nabla^2\right)\Phi+\left(\ddot{\Psi}+\frac{\dot{a}}{a}(\dot{\Psi}-\dot{\Phi})\right)\delta_{ik}\\-\partial_{(i}\left(\dot{\Xi}_{k)}+\frac{\dot{a}}{a}\Xi_{k)}\right)+\left(\ddot{\mathpzc{h}}_{ik}+\frac{\dot{a}}{a}\dot{\mathpzc{h}}_{ik}\right)\Bigg)\delta^{jk}.\label{curvature}
\end{multline}
Thus, a derivative of a delta function in the curvature requires a step function (or worse) discontinuity in the gauge invariant metric variables. 

We are now in a position to analyze how discontinuities could arise. First, it is important to note that the equations \eqref{scalar}-\eqref{svt} giving the scalar, vector, and tensor parts of a tensor $X_{\mu \nu}$ involve solving elliptic and/or algebraic equations, with ``source'' given by components of $X_{\mu \nu}$ and their derivatives. It follows immediately that the scalar, vector, and tensor parts of $X_{\mu \nu}$ are smooth wherever $X_{\mu \nu}$ itself is smooth. In particular, the scalar, vector, and tensor parts of the particle stress-energy \eqref{particleSE} are smooth away from the worldlines of the particles.

Consider, now, the scalar perturbations. Eqs. \eqref{EFE1} and \eqref{EFE2} are elliptic in $\Phi$ and $\Psi$, so these quantities---which fully characterize the scalar part of the metric perturbation---can be singular only where the source terms in these equations are singular. These source terms involve the scalar part of the particle source and the quantity $A$. The scalar part of the particle source is smooth away from the particle world lines. The quantity 
$A$ satisfies the hyperbolic equation \eqref{soundwave}, which, in turn is sourced by the scalar parts of the particle stress-energy. 
We shall analyze the possible singular behavior of $A$ in the Appendix. We shall show there that,
although $A$ can be discontinuous along the sound cone of the source event $q$, there cannot be any discontinuities in $\Phi$ or $\Psi$. Thus, no memory effect can occur in the scalar sector.

Consider, now, the vector perturbations. The quantity $\Xi_i$ satisfies the elliptic equation \eqref{EFE_5} and thus is smooth wherever the sources are smooth. However, the particle source term $\xi_i^{(P)}$ is smooth away from the worldlines of the particles and the fluid source term $W_i$ satisfies the source free evolution equation \eqref{con3} and is thus nonsingular everywhere. Thus, $\Xi_i$ is smooth away from the particle worldlines and no memory effect can occur in the vector sector.

In the next section, we calculate the memory effect occurring in the tensor sector.

\section{The Retarded Gravitational Field and the Memory Effect}\label{sec:retgrav}

The tensor perturbations are described by the quantity $\mathpzc{h}_{ij}$, which satisfies (see \eqref{EFE_7})
\begin{equation}
-\ddot{\mathpzc{h}}_{ij}-2\frac{\dot{a}}{a}\dot{\mathpzc{h}}_{ij}+\nabla^2\mathpzc{h}_{ij} =-16\pi \mathscr{T}_{ij} \, , \label{sw}
\end{equation}
where $\mathscr{T}_{ij}$ is the tensor part of the particle stress energy. Thus, each component of $\mathpzc{h}_{ij}$ in the coordinates \eqref{concoor} satisfies a decoupled scalar wave equation and it suffices to analyze the behavior of solutions to the scalar wave equation. 

We are interested in the contribution to the retarded integral
\begin{equation}
\mathpzc{h}_{ij}(x)=16\pi\int \sqrt{-g(x')}d^4x'G^{\rm ret}(x,x')\mathscr{T}_{ij}(x')\label{FLRWpert}
\end{equation}
arising from a small neighborhood of the source event $q$ (see section~\ref{sec:idealized}), where $G^{\rm ret}(x,x')$ denotes the retarded Green's function for the scalar wave equation 
\begin{equation}
- \ddot{\phi} - 2\frac{\dot{a}}{a} \dot{\phi} + \nabla^2 \phi = -16\pi T \, .
\label{sw1}
\end{equation}
Specifically, we seek to determine whether a discontinuity can arise in $\mathpzc{h}_{ij}$ and, if so, to determine its magnitude. Such discontinuities will give rise to derivative of delta function contributions to the curvature, which, in turn, will produce a memory effect.

To proceed, we need to know the form of $G^{\rm ret}(x,x')$. Consider an equation of the general form
\begin{equation}
L[\phi]=g^{\mu\nu}\partial_\mu\partial_\nu\phi+b^\mu\partial_\mu\phi+c\phi=-16\pi T \, ,\label{PDE}
\end{equation}
where $g_{\mu \nu}$ is a metric of Lorentz signature. It is well known \cite{Garabedian}, \cite{Friedlander} that, in $4$ spacetime dimensions, the retarded Green's function for this equation takes the form
\begin{equation}
G^{\rm ret}(x,x')=\left[U(x,x')\delta(\sigma)+V(x,x')\Theta(-\sigma)\right] \Theta(t-t') \, , \label{Green}
\end{equation}
where $\sigma$ denotes the squared geodesic distance between $x$ and $x'$ in the metric $g_{\mu \nu}$ and $t$ is a global time function. The quantities $U$ (the Van Vleck-Morette determinant) and $V$ are smooth functions; we refer to $U(x,x')\delta(\sigma)$ as the ``direct part'' and $V(x,x')\Theta(-\sigma)$ as the ``tail part'' of $G^{\rm ret}(x,x')$. In general, the form \eqref{Green} for $G^{\rm ret}(x,x')$ will hold only locally in a convex normal neighborhood, but in the case of \eqref{sw1}, the spacetime metric corresponding to \eqref{PDE} is flat, and this form of $G^{\rm ret}(x,x')$ holds globally, with
\begin{equation}
\sigma(x,x')=-(\eta-\eta')^2+(x-x')^2+(y-y')^2+(z-z')^2 \, .
\end{equation}

The Van Vleck-Morette $U$ is determined by integrating an ODE along a geodesic connecting $x$ and $x'$ \cite{Garabedian}, \cite{Friedlander}. The quantities appearing in this ODE depend on $g^{\mu \nu}$ and $b^\mu$ but do not depend on $c$. We could integrate this ODE directly, but we can greatly simplify the calculation of $U$ by working with the rescaled variable ${\tilde \phi} = a\phi$ (where $\phi$ denotes a component of $\mathpzc{h}_{ij}$), which satisfies the equation
\begin{equation}
- \ddot{\tilde \phi} + \nabla^2 {\tilde \phi}  + \frac{\ddot{a}}{a}{\tilde \phi} =-16\pi a T \, .
\label{sw2}
\end{equation}
This change of variables eliminates the term involving $b^\mu$ in \eqref{PDE}, so $\tilde U$ is determined by exactly the same equation for the wave equation in flat spacetime. Thus we obtain the unique solution $\tilde{U}(x,x') = (4\pi)^{-1}$.  However, the retarded Green's function for $\phi$ is related to the retarded Green's function for $\tilde \phi$ by \cite{BHP}-\cite{Poisson} 
\begin{equation}
G^{\rm ret}(x,x')=\frac{a(\eta')}{a(\eta)}\tilde{G}^{\rm ret}(x,x') \, .
\end{equation}
Thus, we obtain
\begin{equation}
U(x,x')=\frac{1}{4\pi}\frac{a(\eta')}{a(\eta)} \, .
\label{U}
\end{equation}
This holds for any FLRW universe, i.e., we have not assumed any particular equation of state $\bar{P}= \bar{P}(\bar{\rho})$ (and, thus, we have not assumed any particular expansion law) in the background spacetime. By contrast, $V(x,x')$ will depend on the expansion history of the FLRW universe between $\eta'$ and $\eta$. Note, however, that by spatial Euclidean invariance, $V$ depends on $(x,x')$ only via $\eta$, $\eta'$, and $|\mathbf{x} - \mathbf{x}'|$.

As previously stated, we are interested in the possible discontinuities in $\mathpzc{h}_{ij}$ resulting from the source behavior near $q$, where we take $q$ to have coordinates $\mathbf{x} = t =0$. To analyze this, let us first introduce a new toy mathematical problem, wherein we consider retarded solutions to the equation
\begin{equation}
-\ddot{H}_{ij}-2\frac{\dot{a}}{a}\dot{H}_{ij}+\nabla^2H_{ij} =-16\pi T^{(P)}_{ij} \, .
\label{modified}
\end{equation}
Eq. \eqref{modified} differs from the equation of interest \eqref{sw} in that we have not taken the tensor part of the particle source and, correspondingly, we do not require $H_{ij}$ to be transverse or traceless.
By \eqref{totalT}, $T^{(P)}_{ij}$
consists of a sum of terms, each one of which has the form $ \delta^{(3)}\left(\mathbf{x} - \mathbf{z}(t)\right) \Theta(\pm t)$, where $\mathbf{z}(t)$ describes a timelike or null geodesic. The contribution of the tail part, $V(x,x')\Theta(-\sigma)$, of $G^{\rm ret}(x,x')$ to the retarded integral is thus a sum of terms of the form
\begin{gather}
H^{\rm tail}_{ij} (x) =\int d^4x' f_{ij}(x') V(x,x')\Theta\left(-\sigma(x,x')\right) \delta^{(3)}\left(\mathbf{x}' - \mathbf{z}(t')\right) \Theta(\pm t') \, ,
\label{Vint}
\end{gather}
where $f_{ij}$ is smooth.
It is not difficult to see that $H^{\rm tail}_{ij}$ is smooth whenever $x$ does not lie on the future light cone of $q$. On the other hand, if $\mathbf{z}(t)$ is a null geodesic and if $x$ lies on (the continuation of) this null geodesic---i.e., if one of the ingoing or outgoing null particles is ``aimed'' directly at an observer at $x$---then the singularities of $\delta^{(3)}\left(\mathbf{x}' - \mathbf{z}(t')\right)$ and $\Theta\left(-\sigma(x,x')\right)$ will coincide and $H^{\rm tail}_{ij}$ will, in general, be ``highly singular'' at $x$ in the sense that, in general, it will be defined only distributionally in a neighborhood of $x$. We exclude such special points from consideration. The case of main interest is one where $x$ lies near the future light cone of $q$ but does not lie on the special direction defined by $\mathbf{z}(t)$ (if null). Then the  $\delta^{(3)}\left(\mathbf{x}' - \mathbf{z}(t')\right)$ singularity in the integral will be transverse to the step function singularity of $\Theta\left(-\sigma(x,x')\right)$ as well as to that of $\Theta(\pm t)$. One can then integrate over $\mathbf{x}'$, leaving one with an integral only over $t'$. The integrand will be proportional to $V(x; \mathbf{z}(t'), t') \Theta(u-t') \Theta(\pm t')$, where $u = t-|\mathbf{x}|$ denotes the retarded time of $x$. The integral over $t'$ thus yields a result of the form $F_{ij}(x) u \Theta(u)$, where $F_{ij}$ is smooth. Thus, we see that $H^{\rm tail}_{ij}$ is continuous (although not continuously differentiable) at $x$. 

The analysis of the contribution, $H^{\rm dir}_{ij}$, of the ``direct part,'' $U(x,x') \delta(\sigma)$, of $G^{\rm ret}(x,x')$ to $H_{ij}$
is similar, with $U(x,x') \delta(\sigma)$ replacing $V(x,x')\Theta(-\sigma)$ in \eqref{Vint}. Again, $H^{\rm dir}_{ij}$ is smooth whenever $x$ does not lie on the future light cone of $q$, and is highly singular if $\mathbf{z}(t)$ is a null geodesic and if $x$ lies on (the continuation of) this null geodesic. If we exclude such special points, then the integral over $\mathbf{x}'$ can again be done, and we are again left with an integral over $t'$. However, now the integrand is proportional to $\delta(u-t') \Theta(\pm t')$, where $u$ is the retarded time of $x$. Consequently, $H^{\rm dir}_{ij}$ has the form $\tilde{F}_{ij}(x) \Theta(u)$ for some smooth $\tilde{F}_{ij}$, and thus it has a discontinuity along the future light cone of $q$.

The above analysis is for the toy problem \eqref{modified}, where we did not take the tensor part, $\mathscr{T}_{ij}$, of the source $T^{(P)}_{ij}$. It would be cumbersome to compute $\mathscr{T}_{ij}$ and then perform a similar direct analysis of the behavior of the retarded Green's function integral involving $\mathscr{T}_{ij}$. Fortunately, we can bypass this by noting that the operation of ``taking the tensor part" commutes with the wave operator appearing in \eqref{sw} and \eqref{modified}. It follows that 
the desired quantity, $\mathpzc{h}_{ij}$, given by \eqref{FLRWpert}, is related to $H_{ij}$ by
\begin{equation}
\mathpzc{h}_{ij}=[H_{ij}]^T \, ,
\end{equation}
where ``$[X_{ij}]^T$'' denotes the operation of taking the ``tensor part'' of $X_{ij}$, as given by \eqref{svt}. Thus, our analysis reduces to extracting information about how the operation of taking the tensor part of a quantity affects the nature of its singularities. To analyze this, we note first that since ``taking the tensor part'' consists of algebraic operations involving differentiations and inversions of Laplacians (see \eqref{decomp}-\eqref{svt}), the tensor part, $\mathscr{X}_{ij}$, of a distribution $X_{ij}$ must be smooth wherever 
$X_{ij}$ is smooth. It then follows that the singular behavior of the tensor part of $X_{ij}$ at $x$ is the same as that of the tensor part of $\psi X_{ij}$, where $\psi$ is any smooth function of compact support with $\psi = 1$ in a neighborhood of $x$. (Proof: $X_{ij} - \psi X_{ij}$ vanishes in a neighborhood of $x$ and hence is smooth there, so the tensor part of this difference is smooth in a neighborhood of $x$.) However, the singular behavior of $\psi X_{ij}$ is characterized by the decay (or lack thereof) of its Fourier transform at large $k_\mu$. The key point is that in the operation of 
``taking the tensor part,''  there are exactly as many total ``inverse derivatives'' from Laplacian inversions in \eqref{svt} as there are differentiations. It follows that the Fourier transform of the tensor part of $\psi X_{ij}$ is related to the Fourier transform of $\psi X_{ij}$ by a function that is everywhere bounded in $k_\mu$. In particular, the decay of the Fourier transform of the tensor part of $\psi X_{ij}$ at large $k_\mu$ cannot be slower than that of $\psi X_{ij}$.

The above argument can be applied to the present case as follows to get the key conclusion that we need. It is easily seen from the explicit behavior of $H^{\rm tail}_{ij}$ found above that the Fourier transform of $\psi H^{\rm tail}_{ij}$ lies in $L^1$ for any smooth function $\psi$ of compact support. Therefore, the Fourier transform of the tensor part of $\psi H^{\rm tail}_{ij}$---which differs from the Fourier transform of $\psi H^{\rm tail}_{ij}$ by a bounded function of $k_\mu$---also lies in $L^1$. But that implies that the tensor part of $\psi H^{\rm tail}_{ij}$ is continuous for all $\psi$, which implies that the tensor part of $H^{\rm tail}_{ij}$ is continuous. {\em Thus, we have shown that the tail contribution to $\mathpzc{h}_{ij}$ is continuous and thus cannot contribute to the memory effect.}

The above conclusion is all that is needed to derive our results on the memory effect, because, as we have seen above, the direct contribution to the retarded solution is universal, and does not depend on the expansion history. Furthermore, Minkowski spacetime lies within the class of $k=0$ FLRW spacetimes to which our analysis applies. Thus, we can relate the memory effect in an arbitrary FLRW spacetime to that in Minkowski spacetime as follows. Consider a source event at $q$ in the FLRW spacetime that is observed at event $p$. Let $\eta_s$ and $\eta_o$ denote the conformal times of the events $q$ and $p$ respectively.
For convenience, rescale the coordinates, if necessary, so that $a(\eta_s) = 1$. This corresponds to choosing the comoving coordinates to corrspond to proper distances at $\eta = \eta_s$. Now, identify the FLRW spacetime with Minkowski spacetime by identifying the coordinates \eqref{concoor} of the FLRW spacetime with global inertial coordinates of Minkowski spacetime.
Place a source and observer at the events ${\bar q}$ and ${\bar p}$ of Minkowski spacetime that are identified in this manner with events $q$ and $p$ in the FLRW spacetime.
Since $a(\eta_s) =1$, the Minkowski source will physically correspond to the FLRW source provided that the masses and $4$-velocities of each of the particles agree (under this identification) at $q$. It follows immediately from \eqref{U} that the direct part, $\mathpzc{h}_{ij}^{\rm dir}$, of $\mathpzc{h}_{ij}(x)$ near $p$ will be a factor of $1/a(\eta_o)$ times the same function of $x^\mu$ as it is in the Minkowski case, i.e., near $p$,
\begin{equation}
\mathpzc{h}_{ij}^{\rm dir} (x^\mu) = \frac{1}{a(\eta_o)} {\bar {\mathpzc h}}_{ij}^{\rm dir} (x^\mu) \, .
\end{equation}
It then follows immediately from \eqref{curvature} that the direct parts of the linearized Riemann curvature tensor are similarly related, i.e., near $p$
\begin{equation}
\delta {R^{\rm dir}_{i00}}^{j} (x^\mu) = \frac{1}{a(\eta_o)} \delta {\bar R}^{\rm dir}_{i00}{}^j (x^\mu) \, .
\label{curvcomp}
\end{equation}

Suppose, now, that we place a gravitational wave detector at $p$, composed of two nearby particles initially at rest in the cosmic reference frame. By the geodesic deviation equation, the deviation vector, $D^i$, describing the displacement of the particles will satisfy
\begin{equation}
v^b \nabla_b (v^c\nabla_c {D}^a) = R_{def}^{\quad a}{D}^d v^e v^f \, ,
\end{equation}
where $v^a$ is the unit tangent to the geodesic. Since $v^\mu \approx 1/a(\eta_o) (\partial/\partial \eta)^\mu$ and the Hubble expansion is negligible over the relevant timescale, we can rewrite this equation as
\begin{equation}
\frac{d^2}{d \eta^2} D^j = R_{i00}{}^j D^i \, .
\label{gd}
\end{equation}
Let $\Delta D^i$ denote the coordinate components of the ``memory displacement,'' obtained by integrating \eqref{gd} twice with respect to $\eta$.
In view of \eqref{curvcomp} and the fact, proven above, that the ``direct part'' of the Riemann tensor contains the full memory effect, we see that 
\begin{equation}
\Delta D^i = \frac{1}{a(\eta_o)} \overline{\Delta D}^i \, ,
\end{equation}
where $\overline{\Delta D}^i$ denotes the corresponding memory displacement in Minkowski spacetime, assuming that the initial displacement was $\overline{D}^i = D^i$. Thus, the relationship between $\Delta D^i$ and $D^i$ in an arbitrary FLRW spacetime differs from the corresponding Minkowski result by a factor of $1/a(\eta_o)$.

Thus, we have shown that {\em if we identify the FLRW spacetime with Minkowski spacetime via the coordinates \eqref{concoor} in such a way that $a(\eta_s) = 1$, and we place the same physical source at $q$ and the same physical detector at $p$ in both spacetimes, then the memory effect in the FLRW spacetime will be a factor of $1/a(\eta_o) = 1/(1 + z)$ smaller than the corresponding memory effect in Minkowski spacetime.} Note that placing the source at the same proper distance at the time of emission corresponds to placing the source at the same angular diameter distance in both spacetimes.

The above result compares the memory effect in FLRW and Minkowski spacetime when the source and detector are at the same proper distance at the source emission time, i.e., when they are at the same location with $a(\eta_s) = 1$. Since the memory effect in Minkowski spacetime falls off as $1/r$, this result may be reformulated in numerous equivalent ways. In particular, we have
\begin{itemize}

\item {\em If the source and detector are placed so that they are at the same proper distance at the time of detection (rather than emission), then the memory effect in the FLRW spacetime is identical to the corresponding memory effect in Minkowski spacetime.}

\item {\em If the source and detector are placed so that the source is at the same luminosity distance in both cases, then the memory effect in the FLRW spacetime is larger by a factor of  $(1+z)$ as compared with the corresponding memory effect in Minkowski spacetime.}

\end{itemize}

The above result is in agreement with the results of \cite{BGY}, \cite{Kehagias}, \cite{Chu1}, and \cite{Chu2} in the cases where the results of those references apply.

\bigskip

\noindent
{\bf Acknowledgements}

We wish to thank Lydia Bieri and David Garfinkle for helpful discussions. This research was supported in part by NSF grants PHY 12-02718 and PHY 15-05124 to the University of Chicago.

\begin{appendix}
\section{Acoustic Shock Waves and Metric Continuity}\label{appendix}

In this Appendix, we show that the the scalar sector does not contribute to the memory effect. The gauge-invariant density perturbation $A$ satisfies \eqref{soundwave}. 
Equation \eqref{soundwave} is a hyperbolic wave equation of the general form \eqref{PDE}, with the Lorentz metric $g_{\mu \nu}$ now being the ``acoustic metric,'' 
\begin{equation}
ds^2= -d \eta^2 + \frac{1}{c_s^2}[dx^2 + dy^2 + dz^2] \, ,
\label{acoustic}
\end{equation}
and with the source term $T$ proportional to the scalar particle fields $\varphi^{(P)}$ and $\psi^{(P)}$. Thus, the retarded Green's function for \eqref{soundwave} takes the general Hadamard form \eqref{Green}, with $\sigma$ replaced by the squared geodesic distance, $\sigma_s$, in the acoustic metric \eqref{acoustic}, i.e., we have
\begin{equation}
G_s(x,x')=\left[U_s(x,x')\delta(\sigma_s)+V_s(x,x')\Theta(-\sigma_s) \right] \Theta(\eta-\eta') \, ,
\end{equation}
where $U_s$ and $V_s$ are again smooth functions in both $x$ and $x'$
Furthermore, it can be seen from \eqref{particleSE} that both $\varphi^{(P)}$ and $\psi^{(P)}$ are obtained from $T^{(P)}_{\mu\nu}$ by algebraic operations (i.e., no differentiations or Laplace inversions). It follows immediately that the source term appearing in \eqref{soundwave} takes the same form (namely, proportional to $\delta^{(3)}\left(\mathbf{x} - \mathbf{z}(\eta)\right) \Theta(\pm \eta)$), as considered in Section \ref{sec:retgrav}. We may therefore repeat the analysis of Section \ref{sec:retgrav} to draw the following conclusion: {\em Suppose that all of the particles in $T^{(P)}_{\mu\nu}$ are moving with velocity smaller than the speed of sound\footnote{If any of the particles are moving with velocity greater than the speed of sound, there will be additional ``Cherenkov radiation'' singularities occurring at points $x$ where the past sound cone of $x$ intersects a particle world line orthogonally (in the sound metric). These additional singularities are not of interest for the memory effect.}. Then $A$ is smooth except on the future sound cone of $q$. Furthermore, on the sound cone, $A$ will, in general, be discontinuous, but it cannot have ``worse'' singular behavior.}

The metric perturbation variables $\Phi$ and $\Psi$ satisfy elliptic equations, with source terms given by $A$ and the scalar parts of the particle sources. It follows that $\Phi$ and $\Psi$ must be smooth everywhere apart from the worldlines of the particles and the points at which $A$ fails to be smooth, i.e., the future sound cone of $q$. We are not interested in the singularities at the particle worldlines. However, on the future sound cone of $q$, $\nabla^2 \Phi$ and $\nabla^2 \Psi$ are at worst discontinuous, so 
$\Phi$ and $\Psi$ themselves are at least $C^1$. Thus, they cannot contribute a derivative of delta-function to the Riemann curvature \eqref{curvature}, and thus do not contribute to any memory effect.

\end{appendix}

\end{document}